# All-optical control of antiferromagnetic domains via an inverse optical magnetoelectric effect

S. Toyoda[1], V. Kocsis[2], Y. Tokunaga[3], I. Kézsmárki[4], Y. Taguchi[1],

T. Arima[1,3], Y. Tokura[1,5], and N. Ogawa[1]

[1] RIKEN Center for Emergent Matter Science (CEMS), Wako, Japan

[2] Institut für Festkörperforschung, Dresden, Germany

[3] Department of Advanced Materials Science, The University of Tokyo, Kashiwa, Japan

[4] Center for Electronic Correlations and Magnetism, University of Augsburg, Augsburg, Germany

[5] Tokyo College and department of Applied Physics, The University of Tokyo, Tokyo, Japan

E-mail: shingo.toyoda@riken.jp

**Antiferromagnets are a promising platform for next-generation spintronics due to their ultrafast spin dynamics and robustness to external fields[1,2]. All-optical control of antiferromagnetic order is essential to fully exploit their potential in energy-efficient and high-speed spintronic and memory applications[3–7]. However, optical writing of antiferromagnetic domains remains a fundamental challenge, as conventional magneto-optical techniques rely on net magnetization, which is absent in antiferromagnets. In certain multiferroic antiferromagnets, the magnetic toroidal moment provides an additional degree of freedom through its inherent magnetoelectric coupling[8–10]. This coupling at higher frequencies results in the optical magnetoelectric effect (OME), which manifests as a directional asymmetry in light propagation and enables optical probing of antiferromagnetic states[11–15]. Here, we demonstrate all-optical writing of antiferromagnetic domains using the inverse optical magnetoelectric effect (IOME) in ferrotoroidic $LiNiPO_4$. The writing process is nonvolatile, non-thermal, and deterministic, driven solely by reversing the light propagation direction. This directional control arises from a strong coupling between the photon linear momentum and the magnetic toroidal moment, enabling the repeatable switching between time-reversed domains with arbitrary light polarization. Our findings establish IOME as a distinct mechanism for manipulating antiferromagnetic order, opening a new paradigm in opto-magnetism driven by photon momentum.**

Modern semiconductor-based information technology relies on two core technologies: optical communication, which enables ultrafast long-distance data transfer, and magnetic data storage, which provides the nonvolatility and essential robustness for memory. Bridging these domains, all-optical control of magnetism has emerged as a promising strategy for realizing next-generation opto-spintronic devices that integrate the speed of photonics with the memory functionality of magnetism[16,17]. Significant progress has been made in controlling magnetization using light, particularly via the inverse Faraday effect (IFE) in ferro- and ferrimagnetic materials, where circularly polarized light induces an effective magnetic field through angular momentum transfer[18–20]. This enables ultrafast and deterministic switching of magnetization without an external magnetic field[17,21–23] (Fig. 1a-c).

Recent growing demand for faster and more energy-efficient computing has shifted the focus of spintronics from ferromagnets to antiferromagnets, which offer inherently faster spin dynamics and immunity to stray fields[1–3]. A major challenge, however, lies in the absence of net magnetization in antiferromagnets, which makes conventional magnetization switching mechanisms ineffective. Overcoming this fundamental limitation calls for developing alternative approaches to achieve reliable and efficient switching in antiferromagnets[3]. A potential strategy for achieving the optical switching of antiferromagnets is to utilize the coupling between light and hidden magnetic multipole orders. Examples of such couplings include Kerr rotation in magnetic octupole systems[24], magnetic circular dichroism in magnetic monopole and axion insulator systems[6,25], and optical magnetoelectric (OME) effect, e.g., in magnetic ferrotoroidic systems[13,14,26–32]. These optical responses can appear even without net magnetization, as they are driven by magnetic multipole moments other than dipole. Among them, ferrotoroidic materials are of particular interest due to their strong and robust coupling with electromagnetic fields. The magnetic toroidal moment, defined as $T = 1/2 \sum_i r_i \times S_i$, where $r_i$ is the position vector and $S_i$ is the spin of the *i*-th site, arises from vortex-like spin arrangement (Fig. 1d). The toroidal moment is odd under both time-reversal (*T*) and spatial-inversion *(P)*, in contrast to magnetization (*T*-odd, *P*-even). This dual-odd symmetry allows the toroidal moment to couple to the photon linear momentum, which transforms in the same fashion, giving rise to direction-dependent optical responses known as the OME effect[11,12,14,33]. This effect manifests as a difference in the complex refractive index between forward and backward propagations along the macroscopic toroidal moment (Fig. 1e). The OME effect has been observed in various ferrotoroidic systems, including $GaFeO_3$ [34], $CuB_2O_4$ [26], and $LiCoPO_4$ [31] with some cases exhibiting nearly perfect unidirectional transparency[27,29,32], demonstrating the strong light-toroidal coupling. This strong interaction raises the intriguing possibility that the inverse process—the inverse optical magnetoelectric (IOME) effect—could serve as an efficient mechanism for optically controlling the toroidal moment and writing antiferromagnetic states through linear momentum transfer of photons.

In this study, we experimentally demonstrate deterministic all-optical writing of antiferromagnetic domains via IOME effect in the ferrotoroidic LiNiPO$_4$. By irradiating intense femtosecond laser pulses resonant with a *d-d* transition of Ni$^{2+}$ ions, we successfully align the magnetic ferrotoroidic moment and demonstrate all-optical isothermal two-way switching of magnetic domains with laser propagation direction reversal (Fig. 1f). Moreover, this process is both repeatable and deterministic, enabling reliable switching between two ferrotoroidic states with opposite toroidal moments, that is, between two distinct antiferromagnetic domain states. Scanning the laser beam demonstrates local, nonvolatile writing of ferrotoroidic domains with arbitrary shapes. In contrast to conventional polarization-sensitive inverse magneto-optical effects, such as the inverse Faraday and inverse Cotton-Mouton effects, this phenomenon is driven by photon linear momentum, allowing the manipulation of AFM states by light with any polarization state. This unique property opens the door to all-optical magnetic recording devices that operate without polarization optics.

LiNiPO$_4$ crystallizes in the centrosymmetric olivine structure with the orthorhombic space-group *Pnma* (point group *mmm*) and has four formula units within a unit cell[35,36]. Its crystal framework is composed of distorted NiO$_6$ octahedra interconnected by PO$_4$ tetrahedra (Fig. 2a). The material undergoes successive magnetic transitions at $T_{N1}$ = 20.8 K and $T_{N2}$ = 21.7 K[37–39]. Between $T_{N1}$ and $T_{N2}$, it shows an incommensurate antiferromagnetic order. A first-order phase transition at $T_{N1}$ leads to the formation of a low-temperature commensurate antiferromagnetic state, which is characterized by Ni$^{2+}$ spins (3d$^8$, S = 1) aligned predominantly along the *c*-axis with a slight canting towards $\pm$ *a*, as shown in Fig. 2b. This fully compensated spin structure with the magnetic point group *mm´m* breaks both *P* and *T* symmetries while preserving the combined *PT* symmetry, thereby allowing a finite toroidal moment along the *b*-axis without net magnetization[36]. Indeed, the OME effect has recently been observed for the light propagating along the *b*-axis in the collinear antiferromagnetic phase below $T_{NI}$ [13].

Figures 2c and 2d show the OME spectra at 5 K for the light polarized along the $E^\omega \parallel a$ and $E^\omega \parallel c$, respectively. Prior to the measurements, the sample was cooled from 25 K to 5 K under simultaneous application of a magnetic field along the *c*-axis (7 T) and an electric field along the *a*-axis (2.8 kV/cm). The cross-product of the electric and magnetic fields, $\boldsymbol{S}=\boldsymbol{E}\times\boldsymbol{H}$, acts as a conjugate field for the magnetic toroidal moment[10], aligning it parallel or antiparallel to the *b*-axis depending on the sign of $\boldsymbol{S}$ (ME poling). Both fields were removed before the measurements. The upper panels show spectra of absorption coefficient α at 5 K, which exhibit three absorption bands. These bands correspond to *d-d* transitions from the ground state $^3A_{2g}$ to the lowest excited states $^3T_{2g}$ of Ni$^{2+}$ ions in approximately $O_h$ symmetry. Due to the distortion of NiO$_6$ octahedra, the $^3T_{2g}$ level splits into $^3A_1$, $^3B_1$, and $^3A_2$, which correspond to the absorption bands centered around 1300, 1400 and 1670 nm, respectively[13]. The absorption spectra show significant changes depending on the direction of $S$, i.e., the relative orientation of the magnetic toroidal moment and the photon linear momentum. The relative change in absorption $\Delta\alpha/\alpha_0$ reaches nearly 40 % for 1450 nm for $E^\omega \parallel a$ and 30% at 1700 nm for $E^\omega \parallel c$, as shown in the lower panels in Figs. 2c and 2d, respectively.

The pronounced OME signal in LiNiPO$_4$ motivates us to explore its counterpart, IOME. Unless otherwise noted, pump light at 1700 nm polarized along the $c$-axis and probe light at 1450 nm polarized along the $a$-axis, both propagating along the $b$-axis, are employed, as these wavelengths correspond to the maxima of the OME response for the two polarizations (see the dashed lines in Figs. 2c and 2d). The pump and probe beams were focused to spot diameters of approximately 140 μm and 80 μm, respectively. The laser spot size remains nearly constant throughout the sample because the sample thickness ~100 μm is much smaller than the focal depth ~7 mm. The penetration depth of pump wavelength is comparable to the sample thickness, ensuring approximately uniform excitation across the thickness of the sample. The magnetic ferrotoroidic moment is detected via changes in the optical absorption, utilizing the normal OME effect at 1450 nm observable with weak light intensity. We show in Figs. 2e and 2f the temperature dependence of the absorption coefficient for the probe light propagating along the $b$-axis. Prior to the measurement, the sample was cooled down from 25 K to 4 K under the irradiation of the pump light without static ME poling. After cooling down to 4 K, the pump light is switched off, and the absorption of the probe light was measured during a warming process. The propagation direction of the pump light during cooling is parallel to that of the probe light (+$k$ pump) in Fig. 2e, while it is antiparallel direction to the probe (-$k$ pump) in Fig. 2f. A pronounced change in the absorption coefficient was observed across $T_{NI}$, indicating the formation of a photo-aligned magnetic ferrotoroidic moment below $T_{NI}$, which persists even after the pump laser is turned off. The optical absorption decreases for the +$k$ pump configuration, while it increases for the -$k$ pump configuration in the low temperature antiferromagnetic phase, signifying that the macroscopic toroidal moment aligns in a direction depending on the pump-light propagation direction. These results provide experimental evidence of IOME that enables deterministic control of the magnetic toroidal moment through the laser poling process. To further validate this effect, we measure temperature dependence of the optical absorption following the ME poling under positive and negative $\boldsymbol{E} \times \boldsymbol{H}$ fields, as shown in Figs. 2g and 2h, respectively. The resulting absorption profiles closely match those obtained via laser poling, indicating that IOME-based laser poling is efficient enough to create single-domain ferrotoroidic states. We note that the effective field directions, $\boldsymbol{k}^{\omega} \parallel \boldsymbol{E}^{\omega} \times \boldsymbol{H}^{\omega}$ in the laser poling case and $\boldsymbol{E} \times \boldsymbol{H}$ in the ME poling case, are parallel to the $b$-axis, consistently aligning the macroscopic toroidal moment along this direction.

To evaluate the efficiency of laser poling in aligning the ferrotoroidic moment, we examine the temperature dependence of optical absorption as a function of pump laser power during the laser poling process, as shown in Fig. 3a. In the absence of laser irradiation, the optical absorption remains nearly constant with temperature. This result indicates the formation of a multi-domain ferrotoroidic state below $T_{NI}$, which consists of stacked layers of toroidal domains along the *b*-axis with a typical thickness of ~10 μm[40]. As a result, the measured absorption represents an average over +*T* and -*T* domains, effectively canceling the directional response associated with individual domains. In contrast, the optical absorption shows gradual decrease (increase) with increasing the laser power for the + *k* (- *k*) pump configuration, indicating a progressive alignment of the ferrotoroidic moment. The pump fluence dependence in Fig. 3b exhibits a saturation behavior, indicating that a single-domain ferrotoroidic state is established above this threshold.

Figures 3c and 3d display the temperature dependence of optical absorption measured after the laser poling with various pump polarization states for the + *k* and − *k* incident directions, respectively. Remarkably, the absorption remains unaffected by the polarization state of the pump light, suggesting that the IOME is fundamentally driven by the linear momentum of photons, rather than their angular momentum. This is in stark contrast to conventional inverse magneto-optical effects such as the inverse Faraday and inverse Cotton-Mouton effects, where the aligned magnetization is inherently tied to the polarization of light. We note that laser-heating effects are also polarization-independent and have been demonstrated to effectively switch antiferromagnetic domains in a ferroelectric antiferromagnet via a dipole-dipole interaction for an electric polarization[4]. However, LiNiPO$_4$ possesses neither macroscopic electric polarization nor magnetization due to its *PT* symmetry, ruling out active dipole-dipole interactions and thereby excluding this mechanism.

So far, the toroidal alignment has been achieved via laser poling, where the sample is irradiated with laser light during cooling across the magnetic transition temperature. We now demonstrate that toroidal switching can also occur at 5 K well below the Néel temperature, establishing that the effect can be triggered in an isothermal and purely all-optical manner. Figure 4a shows the temperature dependence of optical absorption following a photoexcitation at 5 K for the + *k* (blue) and – *k* (red) pump propagation directions. The sample is initially zero-field cooled from 25 K without pump light irradiation. The absorption exhibits only weak temperature dependence (gray line), reflecting the formation of a multi-domain toroidal state. The photoexcitation with the pump fluence of ~ 22mJ/cm$^2$ leads to abrupt changes, depending on the direction of the pump light. The insets of Fig.4a show spatially resolved transmission images taken before and after the photoexcitation at 5 K. The photoexcited region appears as a dark (bright) spot for - *k* (+ *k*) pump propagation, revealing the locally aligned ferrotoroidic domain. This change is not associated with laser-induced damage, as the absorption differences disappears above the $T_{NI}$. These results provide strong evidence that the ferrotoroidic moments are aligned by light via the IOME.

We next demonstrate repeatable, nonvolatile two-way switching between opposite magnetic toroidal domains, that is, between antiferromagnetic domains with opposite order parameters. Figure 4b shows the optical absorption averaged over the photoexcited area during repeated irradiation with pump light alternately propagating in the $+k$ and $-k$ directions. The clearly reversible absorption changes demonstrate the robustness and full repeatability of the all-optical switching of the antiferromagnetic state.

Figure 4c presents the optical absorption at 5 K, averaged over the photoexcited area, under pump irradiation with varying power and alternating incident direction. A clear hysteresis loop is observed as the pump power is cycled and the propagation direction of the pump light is reversed. The sample is initially cooled without applying electric, magnetic, or optical fields, resulting in a multi-domain ferrotoroidic state. With increasing pump fluence, the transmittance gradually increases as energetically favorable ferrotoroidic domains expand via domain wall motion, and it saturates once a single-domain ferrotoroidic state is formed. Reversing the light's propagation direction flips the effective field. The transmittance remains unchanged until the nucleation of oppositely oriented domains, followed by an abrupt decrease due to domain wall motion. It then saturates again as a single-domain state with the opposite ferrotoroidic moment is established. This behavior resembles magnetization switching in ferromagnetic materials under an external magnetic field, and demonstrates that optical pumping acts as an effective field for the magnetic toroidal moment.

A series of transmission images tracing the hysteresis loop is shown in Fig. 4d. The initial image shows no contrast, reflecting the multi-domain ferrotoroidic state. As the pump power increases in the + *k* direction, a bright spot appears, indicating the formation of the single-domain ferrotoroidic state within the photoexcited area. This spot remains even after the pump is turned off, confirming the nonvolatile nature of the induced state. By reversing the propagation direction of pump, a dark spot appears in the transmission image, corresponding to the domain of opposite toroidal moment. It is well established that the cross product of static electric and magnetic fields (***E*** × ***H***) acts as a conjugate field for the magnetic toroidal moment[10,41]. Analogously, the observed hysteresis curve demonstrates that ***E**$^\omega$* × ***H**$^\omega$* parallel to the photon momentum ***k**$^\omega$* can serve as a high-frequency counterpart to this static conjugate field. This insight reveals that the photon momentum can directly couple to and manipulate magnetic order in condensed matter system, enabling the control of toroidal moments in a purely optical, non-thermal, and nonreciprocal manner.

As a final step, we demonstrate the precise writing of arbitrarily shaped domain patterns. Figure 4e illustrates the schematic of the laser writing process, where the pump beam was scanned across the sample surface using a motorized mirror system. In Fig. 4f, we present a transmission image taken in the collinear antiferromagnetic phase at 10 K that clearly demonstrates our ability to imprint ferrotoroidic domain structures with high spatial precision. Photoexcited regions irradiated with pump light propagating in the + *k* (- *k*) direction exhibit higher (lower) transmitted light intensity, indicating the successful creation of toroidal domains with opposite orientations. When the sample is imaged from the opposite side, the contrast is reversed, as shown in Fig. 4g. This behavior is in an excellent agreement with our model, confirming that the observed optical contrast originates from the ferrotoroidic moment induced by the IOME and read out via the OME.

To conclude, we have experimentally demonstrated all-optical writing of the antiferromagnetic ferrotoroidic domains via IOME, enabled by strong magnetoelectric coupling at optical frequencies. The propagation direction of the incident pump pulses unambiguously determines the direction of the toroidal moment, providing direct evidence for the role of photon linear momentum in controlling magnetic order. This switching is deterministic, repeatable, and nonvolatile, and it occurs at a telecommunication wavelength. Although the antiferromagnetic domain formed via the IOME is on the order of 100 μm in the present case, it can, in principle, be reduced to the scale of the wavelength of light ~1 μm. Near-field microscopy or plasmonic antennas[42,43], which confine light below the diffraction limit, may enable nanoscale control of antiferromagnetic domains and allow direct conversion of optical network signals into high-density magnetic memory. This spatial limitation could also be addressed by employing shorter-wavelength light, as the OME effect has been demonstrated across a broad spectral range from microwave[44] and THz[29,31] to near-infrared[13,27], visible[26,34,45], and even x-ray[46] frequencies. In addition, the IOME should not be intrinsically restricted to low temperatures, since the OME effect has been observed at room temperature[47]. These features collectively underscore the promise of this mechanism for future opto-spintronic devices, offering a pathway toward ultrafast, low-power, and nonvolatile optical control of antiferromagnetic states.

# Methods

**Sample preparation**

Single crystals of LiNiPO$_4$ were grown by a floating zone method[39,48]. Two single crystals were oriented by using Laue X-ray photographs, and cut into thin plates of thickness 120 μm and 100 μm with the widest (010) faces. The former sample was used in the measurements in Figs. 2e-h and 3, while the latter one was used in the imaging measurements in Figs.2c,d and 4. The sample surfaces were specularly polished to the thickness of ~100 μm by alumina lapping films.

**Optical absorption measurements**

The sample was mounted on a sapphire substrate in a closed cycle variable-temperature cryostat. A split-coil superconducting magnet was used to apply a magnetic field up to 7 T along the *c*-axis. Two electrodes were attached on the sample to apply an electric field along the *a*-axis, with a spacing of 0.7 mm between them. A laser-driven white light source was used for the optical absorption spectroscopy in Figs. 2c and d. The white light was directed onto the sample, and the transmitted beam was spectrally resolved using a spectrometer and detected with an InGaAs photodiode. The light intensity was modulated by a mechanical chopper at 500 Hz, and the transmitted light intensity was measured using a lock-in amplifier synchronized to the chopping frequency. The optical absorption spectra were calculated by comparing the intensity spectra of the transmitted light measured with and without the sample.

**Pump-probe measurements**

The sample was mounted on a copper holder with a hole diameter 1 mm in a closed cycle variable-temperature cryostat. The light source used for our pump-probe measurements is a regeneratively amplified laser, producing 160 fs light pulses at a 500 Hz repetition rate. Two optical parametric amplifiers were used to independently tune the wavelengths of the pump and probe beams. The two beams were directed onto the sample at a relative angle of less than 7°. The pump and probe beams were focused to spot diameters of approximately 140 μm and 80 μm, respectively. The pump beam was turned off during the absorption measurements to avoid laser heating, except for the measurements shown in Figs. 4b, 4c and 4d. The switching between the pump propagation direction was conducted by inserting a mirror into the optical path. The mirror was located on a motorized movable stage. The probe laser fluence was kept below 0.01 mJ/cm$^2$, while the pump fluence is specified in the main text. For optical transmission imaging, monochromatic light at 1450 nm was obtained by passing a laser-driven white light source through an optical bandpass filter. The transmitted images were captured using a 200 mm telephoto lens for projection and an InGaAs camera as the detector.

## Acknowledgments

S. T. is supported by Grant-in-Aid for Scientific Research from JSPS, Japan (Grant No. JP18K14154 and JP20H01867).

## Author contribution

S.T. conceived the project, carried out the measurements, and analyzed data. N.O. supervised the project. V.K. and Y.Tokunaga prepared the sample. S.T. wrote the manuscript in discussions with N.O. All authors discussed the results and commented on the manuscript.

## Competing financial interests

The authors declare no competing financial interests.

## Data availability

The data that support the plots within this paper and other findings of this study are available from the corresponding author upon reasonable request.

## Code availability

The code used in this study is available from the corresponding author upon reasonable request.

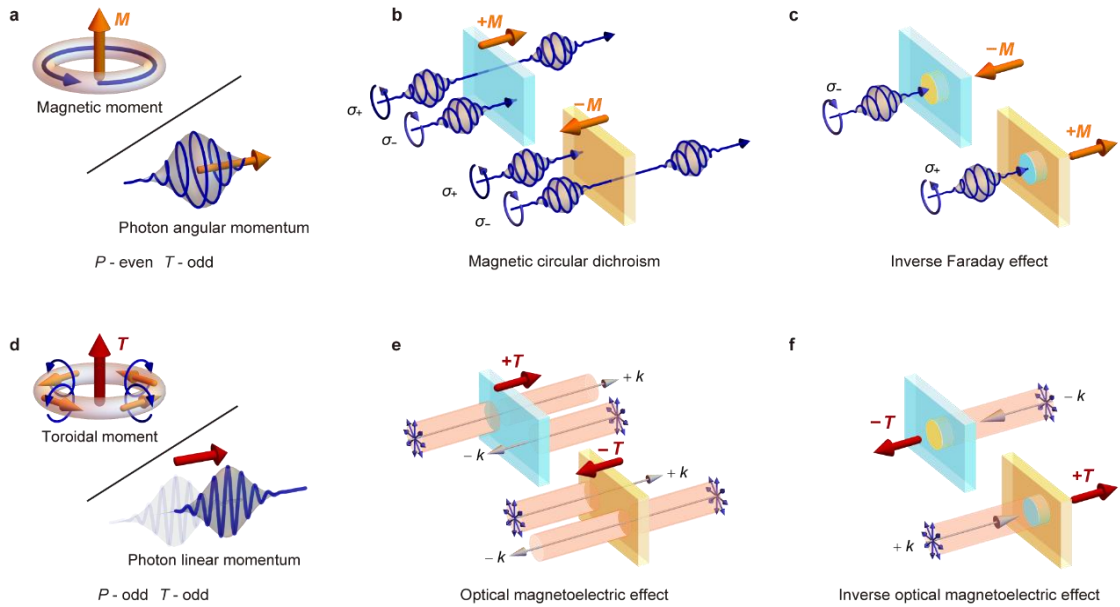

**Figure 1 | Symmetry of magneto-optical effect in ferromagnets and optical magnetoelectric effects in ferrotoroidic materials. a,** Magnetic moment $M$ (orange arrow) formed by a current loop (blue arrow) and circularly polarized light both possess $P$-even and $T$-odd symmetries, leading to helicity-dependent optical responses in ferromagnetic systems. **b,** Magnetic circular dichroism, where optical absorption depends on the helicity of light. The absorption contrast reverses under the reversal of the magnetization direction. **c,** Laser writing of ferromagnetic domains via the inverse Faraday effect. Blue and yellow regions schematically represent ferromagnetic domains with opposite magnetization directions. The magnetization direction in the photoexcited region is determined by the helicity of light. **d,** Magnetic toroidal moment $T$ (red arrow) formed by a spin vortex and photon linear momentum share $P$-odd and $T$-odd symmetries, allowing directionally asymmetric optical responses in ferrotoroidic systems. **e,** Optical magnetoelectric effect, where

optical absorption changes by the reversal of the light propagation direction. The absorption contrast reverses under the reversal of the ferrotoroidic moment. **f,** Laser writing of ferrotoroidic domains via the inverse optical magnetoelectric effect. Blue and yellow regions schematically represent ferrotoroidic domains with opposite toroidal moment. The propagation direction of light (photon linear momentum) determines the direction of the ferrotoroidic moment in the photoexcited region.

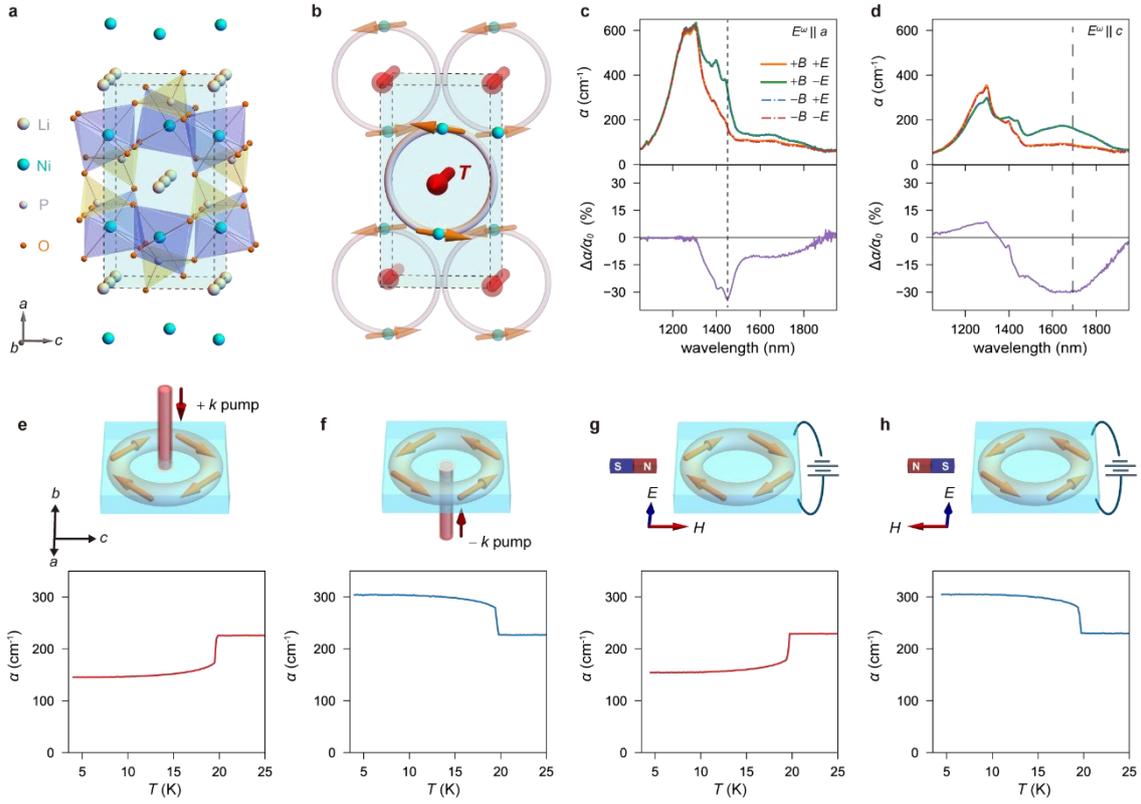

**Figure 2 | Toroidal alignment via laser poling. a and b,** Crystal (a) and magnetic (b) structures of LiNiPO$_4$ in the approximately collinear antiferromagnetic phase projected onto the (010) plane. The orange arrows indicate the spin magnetic moments of Ni$^{2+}$. The rotation of the spin moments produces the finite magnetic toroidal moment along the *b*-axis. **c and d,** Spectra of the optical absorption α (upper panels) and the OME effect $\Delta\alpha/\alpha_0$ (lower panels) for the light polarized along the *a*-axis (c) and *c*-axis (d) measured at 5 K after magnetoelectric poling. The crossed electric field (2.8 kV/cm) and magnetic field (7 T) along the *a*-axis and the *c*-axis, respectively, were applied during the cooling, and removed before the measurements. Here, $\Delta\alpha = \alpha(+B, +E) - \alpha(+B, -E)$ and $\alpha_0 = \alpha(+B, +E) + \alpha(+B, -E)$ denote the difference and sum of the optical absorption between the two opposite toroidal domains, respectively. The dashed lines in c and d represent the

wavelength of probe and pump light pulses, respectively. **e-h,** Temperature dependence of optical absorption at 1450 nm for the light polarized along the *a*-axis. The sample was cooled with the irradiation of the pump light (23 mJ/cm$^2$) propagating in the $+ k$ (e) and $- k$ (f) directions without external electric or magnetic fields. In contrast, the sample was cooled under an external crossed electric (2.8 kV/cm along the *a*-axis) and magnetic (7 T along the *c*-axis) fields without the pump light irradiation, where the toroidal poling field $E \times H$ in g and h was applied in the opposite directions by flipping the external magnetic field. The measurements were performed during a warming process in the absence of external electric, magnetic, or optical fields, confirming the non-volatile nature of the ferrotoroidic order.

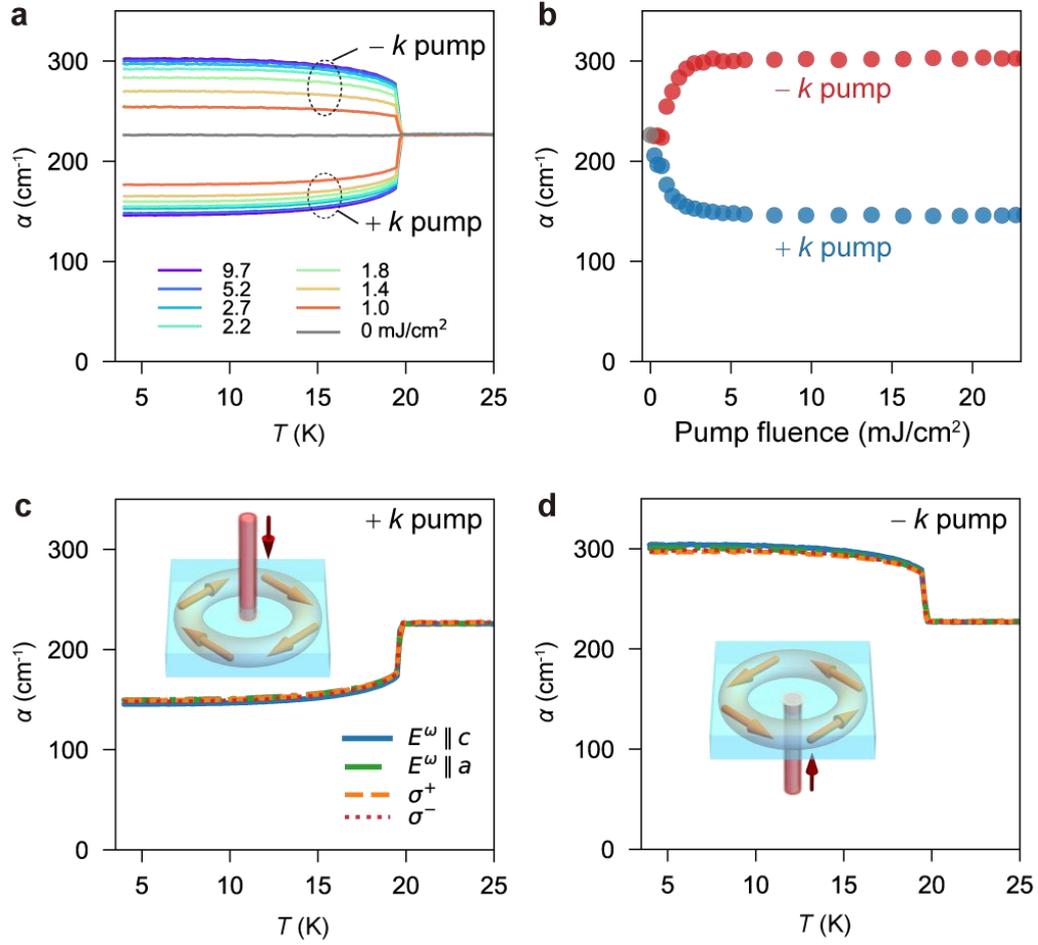

**Figure 3 | Pump power and polarization dependence of the optical absorption. a,** Temperature dependence of optical absorption for the light polarized along the $E^\omega \parallel a$ at 1450 nm measured after the laser poling with various pump fluences propagating along the *b*-axis. **b,** The optical absorption at 5 K as a function of the pump fluence during the laser poling. Red and blue dots indicate the optical absorption measured after the laser poling of the + *k* and – *k* pump propagation directions, respectively. **c and d,** Temperature dependence of optical absorption after the laser poling with several polarization states of the pump light pulses (23 mJ/cm²) which propagates in the + *k* (c) and – *k* (d) directions.

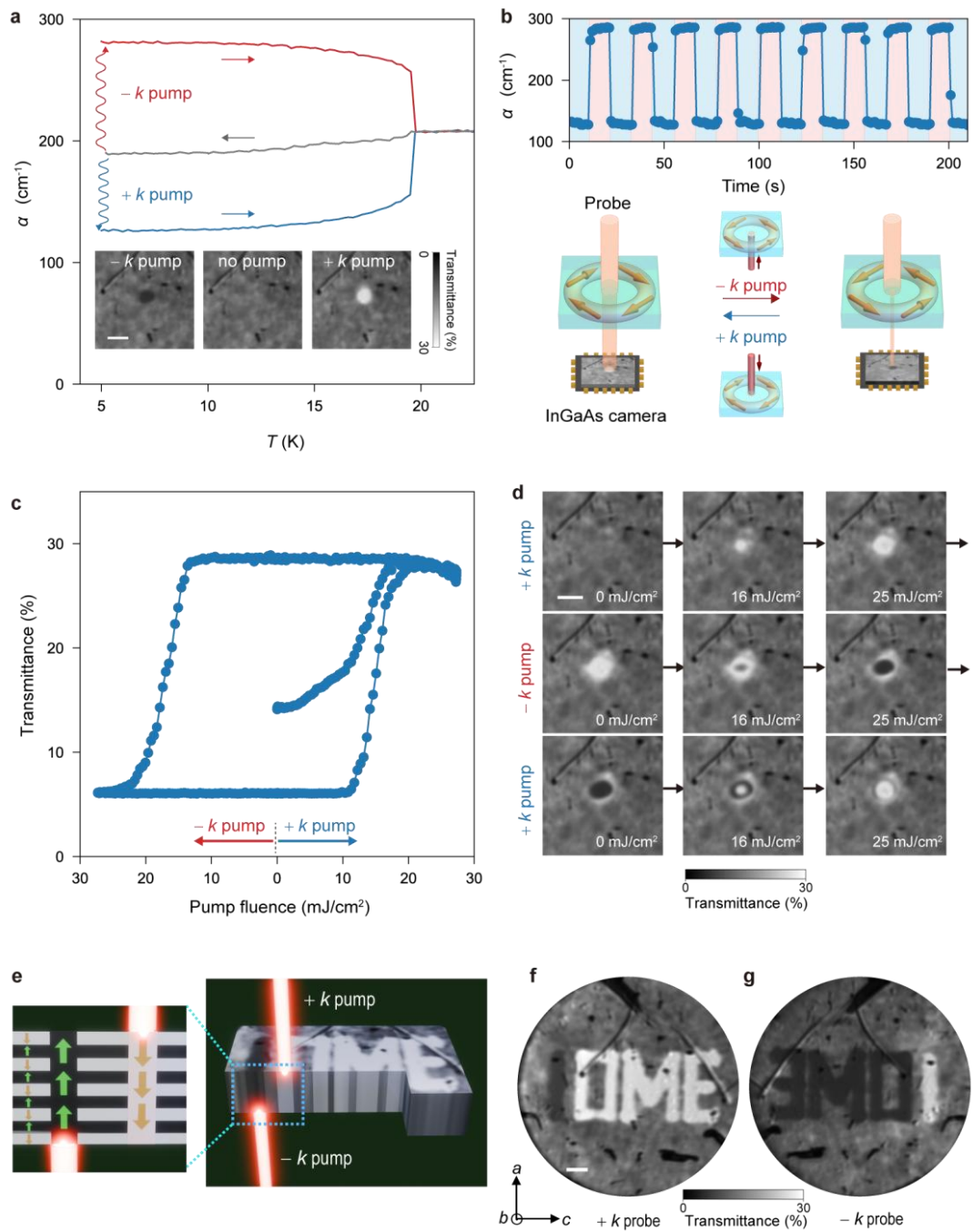

**Figure 4 | All-optical writing of the magnetic toroidal moment. a,** Temperature dependence of optical absorption, which was measured during the zero-laser poling (gray), and warming after the photoexcitation at 5 K for the pump light propagating in the + $k$ (red) and - $k$ (blue) directions with

the pump fluences 22 mJ/cm$^2$. The absorption shows a large change by the photoexcitation, indicating the optical alignment of the macroscopic toroidal moment which changes its direction depending on the incident direction of the pump. **b,** Repeatable switching of the ferrotoroidic moment induced by an alternating irradiation of the pump light (22 mJ/cm$^2$) propagating in the $+k$ and $-k$ directions measured at 5 K. The blue and red shaded regions indicate the durations of $+k$ and $-k$ pump laser irradiation, respectively. **c,** Pump fluence dependence of transmittance for the light polarized along the *a*-axis at 1450 nm averaged over the photoexcited area at 5 K. Across zero, the pump light incidence direction is flipped. **d,** Transmission images that traces the hysteresis loop in c. **e,** A schematic of the controlled laser writing of the magnetic toroidal domains. Flipping the propagation direction of the pump light reverses the orientation of the ferrotoroidic moment. Green and yellow arrows schematically represent the ferrotoroidic domains with the up and down toroidal moments, respectively. The unilluminated region consists of a layered ferrotoroidic multi-domain state stacked along the *b*-axis, with a typical domain thickness of approximately 10 μm[40]. **f,** Optical transmission image obtained at 10 K for the light polarized along the *a*-axis at 1450 nm after laser writing. The region labeled I was pumped from the backside of the sample, whereas the OME regions were irradiated from the front side. **g,** Optical transmission image acquired under the same conditions as in f, but taken from the backside of the sample. Scale bar, 100 μm.